# Few Shot Text - Independent speaker verification using 3D - CNN

Prateek Mishra

Galgotias University, India
Delhi, India
`pr4t333k@gmail.com`

**Abstract.** Facial recognition system is one of the major successes of Artificial intelligence and has been used a lot over the last years. But, images are not the only biometric present: audio is another possible biometric that can be used as an alternative to the existing recognition systems. However, the text-independent audio data is not always available for tasks like speaker verification and also no work has been done in the past for text-independent speaker verification assuming very little training data. Therefore, In this paper, we have proposed a novel method to verify the identity of the claimed speaker using very few training data. To achieve this we are using a Siamese neural network with center loss and speaker bias loss. Experiments conducted on the VoxCeleb1 dataset show that the proposed model accuracy even on training with very few data is near to the state of the art model on text-independent speaker verification.

## 1 Introduction

Every speaker has their own acoustic and behavioral features. This uniqueness in speech data makes it useful for the task like speaker verification. Text-Independent speaker verification is the task of verifying the claimed identity of the speaker where the spoken utterances are different.
In past years, many approaches have been proposed for text-independent speaker verification but they assume the availability of large amounts of training data [1,2]. There has also been work done on speaker verification with few training data [3,4] but they assume that the spoken utterances should be the same in both enrollment and evaluation phase ( Which is referred to as text-dependent speaker verification ). There has been no work done in the past on the task of text-independent speaker verification by assuming very little training data. To this end, we will make use of the siamese neural network as a base network with the raw waveform as input and for learning the feature embeddings we are using 3D-CNN's. Input to the model is very important for making a good verification model and in past, many researchers have experimented their speaker verification model by inputting different forms of audio data like raw waveforms, spectrogram, MFCC features, etc. and among all of those experimented models, those which uses raw waveforms as input have shown comparatively good result. Using Raw waveforms as model input has other advantages as well since they are time-variant audio signals directly collected from the source they require no time and resource for feature extraction. Works by Lee *et al.* and Muckenhirn *et al.* show how raw waveforms on direct input to CNN can be used for audio classification [5,6]. Another important study carried out by Jung *et al.* [2] shows that extraction of multiple frequencies from audio by the kernel of each layer can only be obtained when we process it with a raw waveform.

RawNet proposed by Jung *et al.* [2] is the current state of the art algorithm for text-independent speaker verification but it also assumes a large training data. In this paper, we focus on the text-independent speaker verification assuming that we have little data for training. We have also experimented with our proposed model to achieve near to state of the art performance with very little data so that we can replace the existing face verification system with a text-independent speaker verification system. Our proposed model is the first of its kind that will unravel this problem and thus will provide possibilities for many researchers to implement new procedures by using very little training data.

## 2  Related Works

Research into text-independent speaker verification and few shot speaker verification is present in a huge amount but few shot text-independent speaker verification has received no attention from the signal processing community. Although, there are a few key works that precede this paper.As the proposed methodology uses 3D CNN and few shot learning technique to verify the speaker, the rest of this section analyses works that are related to this methodology.

In the past 3D Convolutional Networks have shown promising improvements in Learning Spatio-temporal Features as investigated by Torfi et al. [1] they are feeding MFEC features to the 3D CNN model and then calculate the similarity score using cross-entropy loss. Despite its improvements over DNN methodology for speaker verification, it lacks in the extraction of spatial and temporal features and calculating similarity score.

Studies carried out on various feature extraction methods and their analysis [11] have shown that Raw waveforms have more Spatio-temporal information than the processed input features which are fed into the model and apart from it raw waveform input model requires less hyperparameter exploration. verification of the speaker is an important step after extracting features and for verifying the speakers accurately the model should learn intra-class and inter-class covariance. And for capturing such information speaker bias loss [12] and center loss [13] can be used.

Few Shot Learning can be defined as the ability of a model to learn from very few labeled training samples. The recent work carried out by Prashant et al. [3] has proposed a deep neural network based model which takes in audio in spectrogram form and then the class similarity is calculated using prototypical loss.

## 3 Methodology

We propose a model consisting of a Siamese neural network for carrying out few-shot learning task and 3D-CNN for learning the feature embeddings which is shown in figure 1 and 2 respectively.

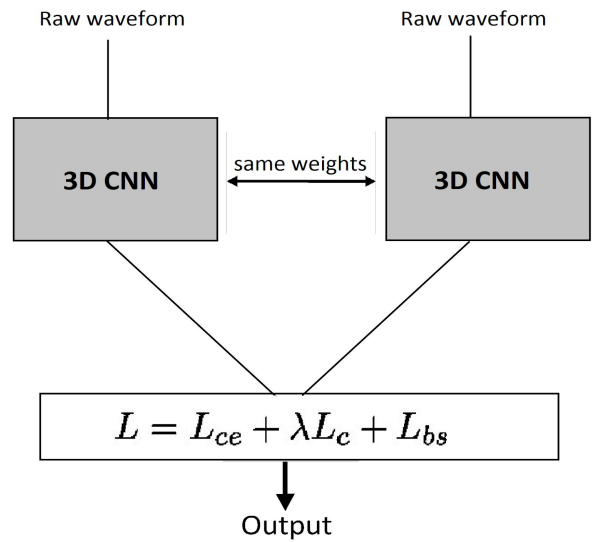

**Fig. 1.** Flow diagram for Few shot speaker verification using Siamese neural network

### 3.1 Dataset Preparation

We are using a subset of VoxCeleb1 dataset [14]. It contains 148,642 utterances of 1,211 speakers which are extracted from 21,819 videos. Although the dataset comes with the list of pairs for verification of two samples but there is no information on the standard split for the few shot learning methodology is given. So for preparing the training and validation data, we split the entire dataset into 70:30 and for every speaker, in each split, we took 5 random audio samples of 3 seconds.

### 3.2 Model architecture

The objective while designing was to make a neural network that can extract the Spatio-temporal features from the raw waveform very well. Figure 2 shows the structure of a NN used for extracting the feature embeddings in our proposed model. It consists of 3D convolution, 3D Max pooling, and Batch Normalization layers stacked together.

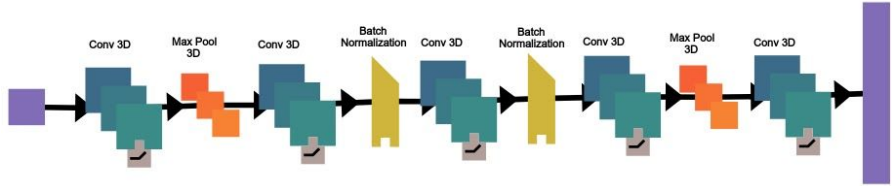

**Fig. 2.** The flow of diagram for the 3D - CNN architecture used for learning feature embeddings.

On experimentation, we observed that 3D-CNN have captured more spatiotemporal features from the utterances then embeddings learned by PLDA [7] and x-vector [8] models.

The Siamese neural network consists of two identical sub neural networks which share the same weights with each other. A Distance-based learning technique is used for training the neural network i.e. for a similar class of input the output distance will be smaller than the distinct class input.

### 3.3 Loss function

The neural network is trained using three different loss functions together [2] which is formulated as in equation 1.

$$L = L_{ce} + \lambda L_c + L_{bs} \quad (1)$$

where $L_c$ is the *center loss,* $L_{ce}$ is the categorical cross-entropy loss, $L_{bs}$ is the speaker-bias loss and $\lambda$ is the hyperparameter for the center loss and has a value of 0.003. We are using Center loss for minimizing intra-class covariance in the learned feature embeddings and Speaker bias loss for maximizing inter-class covariance in the learned feature embedding.

### 3.4 Implementation

For implementing our proposed model we used Keras, a deep learning framework, and trained it on Google Colab which is a free service provided by Google for training neural networks on Cloud. The complete model was trained for 90 epochs with a batch size of 32 at the learning rate of 0.001.

## 4 Results and Discussion

In Figure 3 (a) (b), the training and validation loss, accuracy is shown. The model achieves the final validation accuracy of 0.92 and validation loss of 0.1 on complete training.

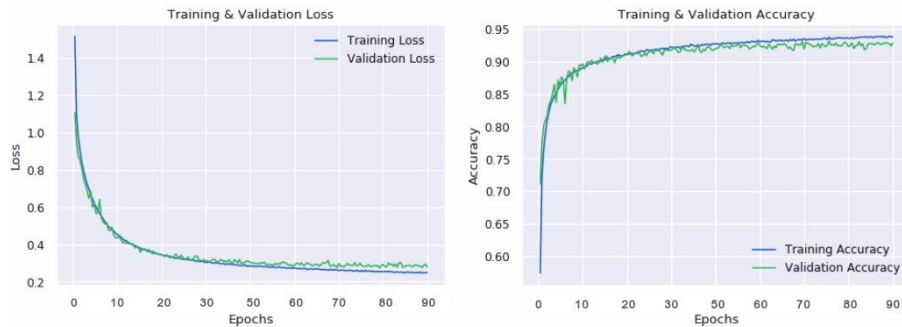

**Fig. 3(a).** Training and Validation loss of the model. **3(b).** Training and Validation accuracy of the model.

Table 1 shows a comparison of equal error rate [ERR] of the works done in the past on text-independent speaker verification and how our model performs.
The RawNet model shows the lowest EER of 4.0% then all the existing models and Our proposed system has an ERR of 6.1% which is near to the RawNet's performance.

**Table 1.** Comparing all the results of speaker verification on the VoxCeleb1 dataset.

| No. | Proposed By | EER % |
|---|---|---|
| 1 | Nagrani *et al.*[9] | 7.8 |
| 2 | Jung *et al.* [10] | 7.7 |
| 3 | Shon *et al.* [8] | 5.9 |
| 4 | Jung *et al*. (RawNet) [2] | 4.0 |
| 5 | **Our Proposed model** | 6.1 |

# 4  Conclusion

In this paper, we propose a novel, simple, and efficient model for text-independent speaker verification assuming very little training data. Various techniques from input to objective function have been explored so that our proposed model achieves near to state of the art model accuracy. The proposed architecture outperformed many architectures proposed in the past and is having an EER of 6.1% which is near to RawNet's EER. One of the main reasons why facial recognition systems are used widely as a biometric is because they require very little data for training. It is expected that the proposed work with accuracy near the state of the art model will become an alternative for the existing recognition system and will uncover the capabilities of audio data and provide several researchers an opportunity to explore by applying new methods.